# Online Decision Process based on Machine Learning Techniques

Tanzila Saba[1,2]

[1]College of Computer and Information Sciences Prince Sultan University Riyadh 11586, Saudi Arabia

[2]School of Computing Universiti Teknologi Malaysia, Malaysia

**Abstract:** This paper analyses role of internet in marketing and its influences on business decision-making process. It explains how the decision maker collect variety of information about customers through internet and analysis this data to better use it in enhancing the processes and the overall performance of the organization. In addition, how each department in an organization collaborates and use these information through data warehousing. Accordingly, a business intelligence model is proposed for web segmentation that divides potential markets or consumers into specific groups and analysis them for better decision making. The model further plans to push the significance of web opportunities in directing the web division and gathering client information. It is exhibited how marketing information system include customers, equipment and procedures analysis contribute to help decision makers make better decision.

## 1. Introduction

For as far back as couple of decades, organizations are seeing the power of quick changes in utilization of data innovations, for the business choice making purposes. Past studies demonstrated that, in the 1990s, organizations used to see the internet for the most part as a specialized instrument - alluding to the email and media abilities, sending or downloading records, and so on. In order to satisfy some essential business capacities, for example, gathering data by investigating other web destinations; giving client booster and directing on line exchanges [1-5]. Consequently, the real discernment alluded to administrations that are not gave in some different courses, for example, by phone or fax, which can make certain astigmatism, leaving the Internet potential uncovered. At the point when dissecting the fresher information in regards to the utilization of the business sector data, it appears that the Internet has still not been completely perceived as a crucial wellspring of showcasing insight. Consequences of the assessment connected on the managing account frameworks demonstrated that there is an open door for abusing the internet as an advertising exploration device, albeit unused because of slant toward the on line exchanges [6-12].

The internet is no more the subject of exploration, but instead a pertinent assessment apparatus. It has turned into the method for enhancing the general overviews and sociology as opposed to depending on the web for insignificant correspondence needs [13-17]. Other than the part of exploration and correspondence, the Web assumes a part of offering and promoting channel upgrading the general intuitiveness of the organizations today, with a specific end goal to stop that potential. This paper surveys significance of Web as an assessment and Business Intelligence instrument. It examines the Web-based division as a way toward building up the business method that empowers comprehension of Web clients' conduct such as target gatherings and subsequently dealing with associations with clients in best conceivable way. A mixed bag of systems and methods in light of the Web and client relationship administration assume a fundamental part in advertising assessment [18-24].

## 1.1 Background

According to the Ivana Kursan about the "Role of the internet in marketing research and business decision-making". Nowadays internet became a popular and the most important tool for searching. Most of the companies now cannot complete their researches and processes without referring to the internet. But the question is how these companies get the information for their marketing and analysis from the internet. Huge number of companies uses social media to gather information about the customers, they collect the data and analyze it to make best decision making. Some of these social media they use are Facebook, Twitter, Instagram and so forth [25-30].

In addition, companies use "CRM" customer relationship management technique to keep in touch with customers through internet. By using this technique, companies can continually develop their services and their decision making.

## 2. Business intelligence

Business intelligence (BI) is an innovation driven procedure for investigating information and showing significant data to help corporate officials, business supervisors and flip side clients settle on more educated business choices. BI envelops an assortment of instruments, applications and techniques that empower associations to gather information from inner frameworks and outer sources, set it up for assessment, create and run questions against the information, and make reports, dashboards and information visualizations to make the investigative results accessible to corporate chiefs and in addition operational specialists. The potential advantages of business intelligence projects incorporate quickening and enhancing decision making; upgrading inner business courses of action; expanding operational effectiveness; driving new incomes; and increasing competitive advantages over business rivals. BI frameworks can likewise help organizations distinguish business sector patterns and spot business issues that need to be tended to [31-35].

BI information can incorporate authentic data, and new information assembled from source frameworks as it is created, empowering BI investigation to backing both vital and strategic decision-making techniques. At first, BI devices were basically utilized by information investigators and other IT experts who ran investigations and delivered reports with inquiry results for business clients. Progressively, be that as it may, business administrators and specialists are utilizing BI programming themselves, because of the advancement of organization toward oneself BI and information disclosure instruments [36-40].

Business intelligence innovations incorporates information visualization programming for outlining outlines and different info graphics, and also devices for building BI dashboards and execution scorecards that show pictured information on business measurements and key execution pointers in a simple to-handle way. BI applications could be purchased independently from distinctive sellers or as a major aspect of a brought together BI stage from a solitary merchant as depicted in Figure 1.

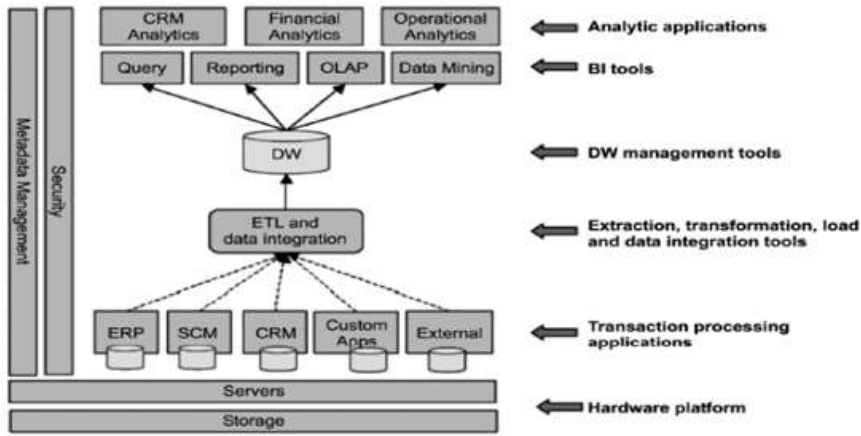

Figure 1: BI Environement

BI projects can perform cutting edge investigation, for example, data mining, prescient assessment, text mining, factual investigation and huge information assessment. Much of the time however, progressed assessment undertakings are led and oversaw by particular groups of information researchers, statisticians, prescient modelers and other talented investigation experts, while BI groups administer clearer questioning and investigation of business information [36-40].

**2.1 Real time application of BI tools in business.**

Business intelligence tools are software programs that are employed to obtain, examine, categorize, refine, and deal with and making reports of information. These tools retrieve the data from the data repository [41-43]. There are various tools that are designed to serve different enterprises' needs. The following list demonstrates three different BI Tools that are commonly used:

- SAP BUSINESSOBJECTS: SAP® Business Objects TM Business Intelligence (BI) is an adjustable an elastic tool that intended to assist the user in a more simply determine the knowledge for enhanced enterprises decision making process [41-43]. Figure 3 illustrate the user interface of SAP BUSINESSOBJECTS platform.

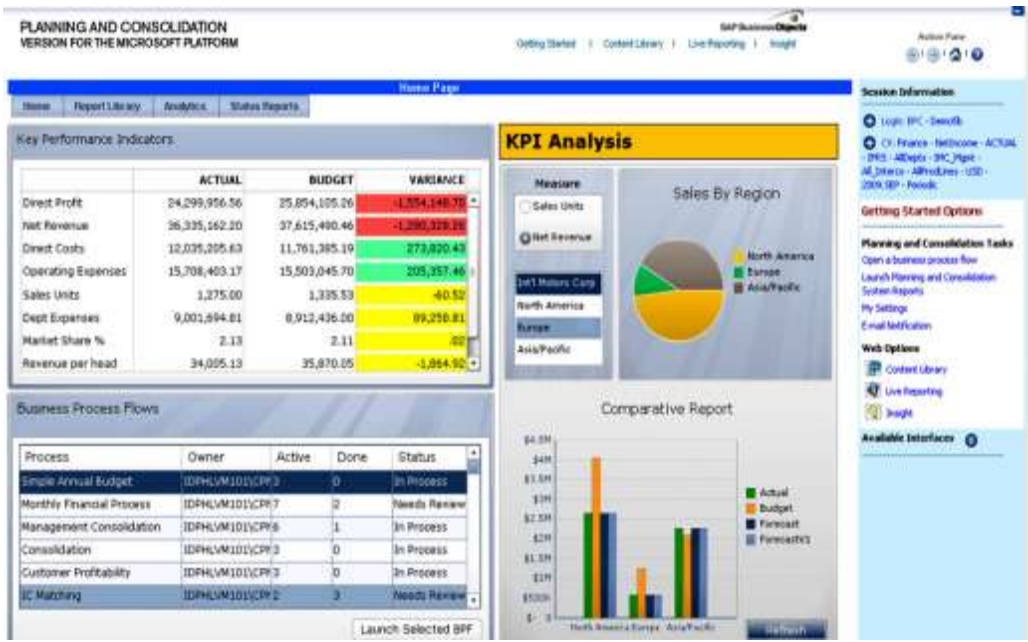

*Figure 2 "Dashboard within SAP BO Planning and Consolidation"*

- Oracle Hyperion Enterprise Performance Management System: is a business intelligence tool. It delivers reports for management and evaluation from data providers, including transactional systems, data repository, SAP BW, Hyperion System 9 BI+ Analytic Services TM, and Hyperion financial applications [44]. Figure 4 illustrates the dashboard of Hyperion.

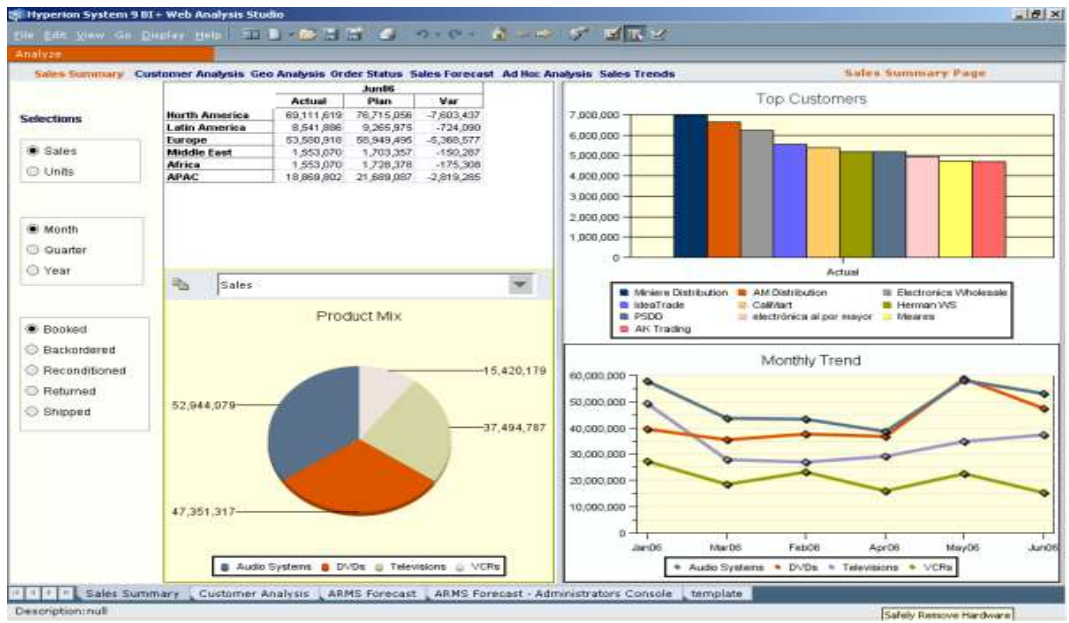

*Figure 3* Hyperion Web Analysis simple to use web browser based dashboards

3. **The Internet as a Research Tool**

*"Information technology (IT) contributed to the growth of world economy. In the network economy, business applications and management must embrace the Internet in order to survive in the e-Commerce age."* [45,46]

From the point of view of building up a productive marketing strategy, the internet gives better insights of knowledge into here and there concealed and distracted information in regards to clients, their effects on business, customer's behavior and buying decisions. It likewise offers an open door for organizations to make a picture, offer data about items and administrations, create associations with gainful clients, better comprehend the purchaser purchasing practices, guarantee ceaseless item enhancements regarding clients' necessities, and so forth. Nevertheless, a few studies demonstrated couple of distinctive observations. In any case, one must look at the internet, as a research tool, to the more customary method for directing statistical surveying.

The internet is a far less expensive and simpler medium for directing research and has various different profits. They incorporate a chance to study a high number of respondents immediately, simplicity of leading an overview in several clicks, cheap respondent achieve (bigger example), prescreened boards (brief reactions to online surveys), or fast turnaround (research and results in a brief time of time). Different studies take note of that the real favorable circumstances of the Web-based research as the potential outcomes for focusing on a bigger populace, adaptability and control over arrangements, basic information passage, high interest, utilization of a mixed bag of media, straightforwardness of organization, and so on. At the point when dissecting information accumulation procedures, it is critical to note that there is a developing pattern of managing the Web-based information gathering approach that has various preferences over other information accumulation approaches.

Accordingly, the internet exploration ought to be utilized alongside other conventional techniques keeping in mind the end goal to cover all client portions. To be specific, some target gatherings are effectively focused by the internet (e.g. more youthful individuals), while others are not reachable along these lines. This makes to the determination that the internet, as a marketing too, achieves its maximum capacity in mix with customary logged off exploration techniques [47].

An accentuation is put on the use of the internet as a manifestation of a propelled assessment device for the better division of the potential clients; however, it might be much more imperative to give continuous information, which can be accomplished by a few innovations. They incorporate, for case, web administrations, being normally characterized as a method for following and checking the business exercises continuously [48].

4. **Marketing Information System based on Business Intelligence Model**

Marketing information system could be defined as according to Kotler and Keller a marketing information system involves: "Individuals, hardware, and methodology to accumulate, sort, examine, assess and appropriate required, convenient and exact data to marketing decision makers."

Information technology plays a significant impact on the business in terms of controlling data, such as storing data, looking for a certain data within an enterprise, and applying data analysis.

## 4.1 Marketing information system components

Marketing information system consists of four significant independent internal units: data bank, measurement-statistics bank, model bank, and communication capabilities. These independent units or modules communicate with two other exterior units: the user and the surrounding conditions that impact the enterprise's marketing actions. The data bank is defined as repository where data can be deposited and must enable alteration and retrieval of data. The manager can view the data in a direct way and a statistical analysis can be applied to it as well. Furthermore, the measurement-statistics bank comprises mechanisms for acquiring and examining considerable decisions that are influenced by individuals. The model bank allocates a heterogeneous selection of marketing models at various extents of difficulties suitable to the comprehensive insight and solution of marketing issues. The last system unit or part is the communications capability. It delivers a two-way communication flow across the user and the system and considered to be a significant part since a valid communication is essential [50].

## 4.2 Sources of data

Internet is considered as a helpful way to manage and carry out primary research and secondary research. Internet provides various tools that are free of charge and provides an ultimate way to attain a great amount of individuals for a cost that remains low. Figure 4.2 illustrates the data sources for both primary research and secondary research.

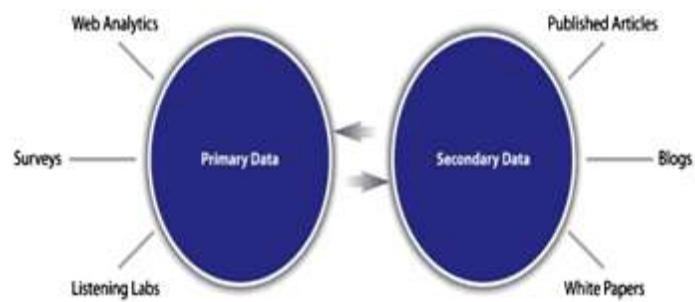

Figure 4 "*Sources of Primary and Secondary Research Data*"

### 4.3 Primary Research

Primary research can be defined to include collecting data for an intended research duty. Primary depends on the data that have never been collected before and the type of data can be qualitative or quantitative. Primary research is beneficial for examining the market and generation hypotheses. In general the qualitative types of data are collected at this level. For instance, using a virtual community to conduct a research to indicate customer's demands that are not achieved before to delivery possible solutions. Additional quantitative research can examine what percentage of customers faces the same problems and consider possible solutions that effectively cover those needs.

### 4.4 Online Research Communities

Despite the fact that online communities are considered as a rich source for secondary research, online communities can also deliver primary data. For instance, Fast Lane is a blog powered by General Motors that helps in collecting research data. The blog can be used as a medium to extract or capture the responses on a specific research problem.

### 4.5 Listening Labs

Testing is a significant phase to ensure that what the web-based application is satisfying the end-user's needs. Listening labs establishing a testing medium in which the use of web-based application by customers can be perceived.

### 4.6 Secondary Research

"Statistical surveying in view of auxiliary assets utilizes information that as of now exists for investigation. This incorporates both inner information and outside information and is helpful for investigating the business sector and promoting issues that exist."[51].

The secondary research should be performed before the primary data research. The secondary research should be deployed in creating the context and the constraints of primary data research. Secondary data can be used to for the following purposes:

1. The data can deliver sufficient information to effectively deal with problems and find the solutions.

2. The secondary data can deliver the material for hypothesis

3. Examining through secondary data is required as an initial step for primary research, as it can deliver information that is related to such as sample size that used for usability testing.

4. To ensure the correctness of primary research.

All the behavior that is done on the website can be stored in the server logs. Another source of data includes the customer interactions with the customer service that is used to research obtaining customer's satisfaction. Social media can also be used to capture customer's opinion towards the products or services [52].

5. **Web Segmentation**

"Segmentation is the process of dividing potential markets or consumers into specific groups. Market research analysis using segmentation is a basic component of any marketing effort. It provides a basis upon which business decision makers maximize profitability by focusing their company's efforts and resources on those market segments most favorable to their goals."[56].

As per creators of "Web promoting exploration: opportunities and issues" specialists can utilize the internet so as to satisfy three different sort of assessment objectives: examine the way in which internet serves marketing needs, they can also employ the internet as a replacement of the traditional surveys, and can be used to examine internet costumer's actions. The research demonstrated the personality categories on the basis of web segmentation and intended to examine the click stream method among three kinds of personal characteristics. The detailed assessment of the click stream assures that a web segmentation can be performed and arrange web users as individuals who disposed to comply, aggressive, and detached [57].

In addition to that, another research was conducted on online-based pharmacies examined the affectivity of marketing procedures and the collected information about costumers on web by examining the user transactional model as an instance of refraining of the web capabilities for the market segmentation. As a result, shows four basic consumer classifications examining the means of decision-making and actions across the web.

The research also indicated that web segmentation is considered as an effective way to perceive an intended analysis about the customer demands for businesses that are online-based [53].

It is unreasonable to create and market a product that contentiously acts in accordance to costumer's demands. Products and services are distinguished from one enterprise to another and could be designed and created according to various specifications and preferences. A significant key of web segmentation resides in the capability of identifying identical segments that appointed to aimed groups in order to implement and assist every single correctly in another word, based on the demands of the individuals who correspond to that segment. The past situation is appropriate for both disconnected from the net and online medium. Overall, the analysts recognized the unmistakable attributes of both the online and disconnected from the net division. The offline segmentation essentially need the segment to be large enough to be measured can be retrieved when needed, interesting, distinct, and not likely to change. On the other hand, the online segmentation depends on the fundamentals of interrelation of measure that also point out the ability to be scaled and the ability to adjust to new conditions. Furthermore, the authors strongly suggest the impact of expand or reduce the number of population. That indicates the availability to be achieved, can be retrieved when needed, able to be scaled, the ability to adjust to new conditions as the significant web segmentation measurable factors and necessary for recognizing and keeping track the complicated and characterized by change web actions of consumer[54,55].

The customer actions can also be retrieved from the highest visit to a certain product or a brand, time consumed, installation, the way that a page is used, etc. Furthermore, costumer's actions or behavior that has no fixed pattern can be viewed from the customer's rate of usage. All activities that occurred before, such as transactions, costumer's products payments can also be retrieved discovered and costumer's favorite choices can be examined according to the search specification on the website. For that reason, the rate of visit to a specific web site can recommend for the enterprise to include remarkable offers on some specific pages, to simplify the exploration of the product in which the user is looking for. In case of data collection, there are three main techniques: the observation, survey, and applying experimentation. Observation can be performed with the assistance of a researcher or a customer sample of information. This technique is used to gather quantitative data, such as the amount of advertisements that are located in a single page. Moreover, observation can also be used to gather qualitative data such as measuring the level of customer's satisfaction towards the current provided services or the customer's experience toward the web page in general, such as user interface, website component etc. In this circumstance, Cox and Dale brought up the discriminating perspectives that check the site execution and the great general experience of site regarding web substance. Counting the administrations and item that are given by the site, the structure and the way how association between the costumer and the site affect the span that the client spent to explore the site, and in addition the unwavering quality that the site accommodate the costumer [56].

Questionnaire or feedback forms can be characterized into site study, overview that is sent by means of email, content based review frame that can be recovered from a certain site. Investigative study, for example, web research component by leading of test site to increase client's activities towards the modifications that are connected on the site. It's conceivable to perform an adjustment regarding the site graphical client interface, references or the route in which the client can see the site. Specialist can solicit a number from individuals who utilizes the web to participate in the site experimentation i.e. to explore through the site to give their own particular criticism on some alluded perspectives. Over a few specialists, the proficiency of the led investigation could be seen, who analyzed the activity of the site clients by applying the "succession arrangement technique", which contained different components that are obligated to change, for example, website page segment, items, and offers. The acknowledgment of sections is subject to the conduct structure. The figure underneath outlines the relationship between the internet-based exploration segments.

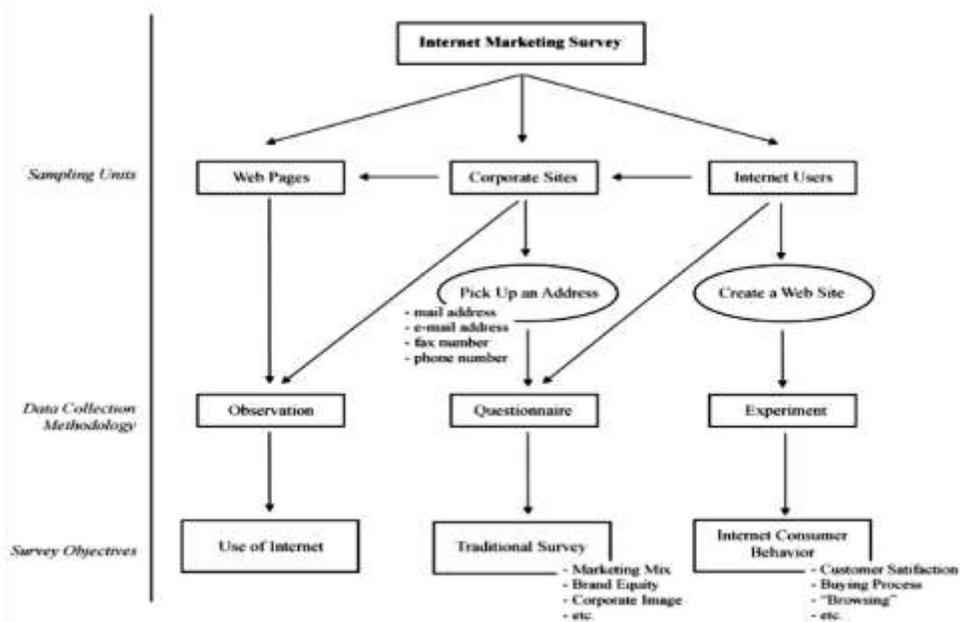

**Figure 5**. Internet Marketing Research Model

6. **Discussion and Analysis**

A survey was conducted online to determine the respondent's reactions towards the marketing research surveys in terms of their willingness to share their personal information. The survey asks for age, name, e-mail, gender and what type of personal information they are willing to share with the organization to support them constructing marketing research. The respondents were asked about what motivates them to participate in surveys and actions like for money, to help, curiosity, to enter prize draws or just because they believe that their feedback and opinion is important. The Figure 6 illustrates the respondents' feedback [18].

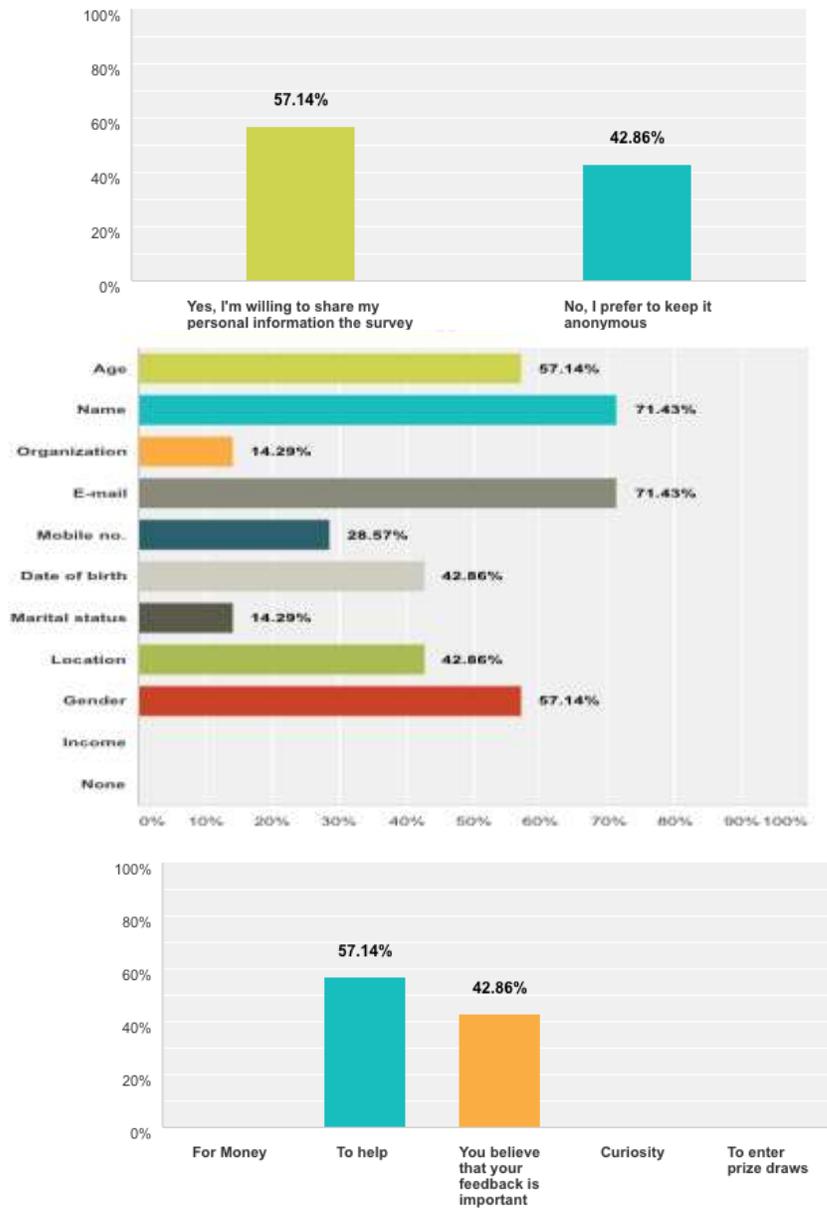

**Figure 6:** Data Analytics

### 7. Conclusion

The paper has assessed web in marketing and regular business decision making. It's focused on the presence of different information sorts, distinctive information sources, additionally various routines and strategies of information gathering (that could be valuable when directing the Web research) are explored. Exceptional accentuation is to dig out significance of Web division that empowers the recognizable proof of homogeneous portions and conveyance of required data to certain target bunches. From this point of view, there are a few pragmatic uses of diverse Business Intelligence apparatuses and innovations in promoting. Concerning customary strategies for business division, supervisors ought to utilize the web division opportunities, so as to characterize and comprehend the web clients, their inspiration and purchasing propensities. It is insufficient to utilize general information or basic reports to completely comprehend the online clients and adjust the advertising technique (showcasing blend) successfully. Advertisers/administrators ought to pay consideration on the corporate Web substance, structure and route in the event that they need to market the organizations in the best conceivable way. It is critical to direct marketing research on the web and to execute web assessment or different routines as often as possible, as to gather the ongoing information. Finally, it is critical to urge organizations to utilize the web as an intelligent medium for creating an association with their clients. With the development of the internet and innovation advancements, it is important to be exceptional with such advances, so as to exploit the maximum capacity that the web offers.

**Acknowledgement**

This work was supported by Artificial Intelligence and Data Analytics (AIDA) Lab Prince Sultan University Riyadh Saudi Arabia